\documentclass[10pt, conference, letterpaper]{IEEEtran}
\IEEEoverridecommandlockouts
\usepackage{cite}
\usepackage{amsmath,amssymb,amsfonts}
\usepackage{graphicx}
\usepackage{textcomp}
\usepackage{xcolor}
\usepackage [english]{babel}
\usepackage [autostyle, english = american]{csquotes}
\MakeOuterQuote{"}
\usepackage{enumitem}
\usepackage{graphicx}
\usepackage{float}
\usepackage{dsfont}
\usepackage{dsfont}
\usepackage{graphics} % for pdf, bitmapped graphics files
\usepackage{epsfig} 
\usepackage{tabularx}
\usepackage{placeins}
\usepackage{multicol}
\usepackage{multirow}
\usepackage{tabularx}
\usepackage{pdflscape}
\usepackage{bbm}
\usepackage{bbding}
\usepackage{subfigure}
\usepackage{wasysym}
\usepackage{amssymb}% http://ctan.org/pkg/amssymb
\usepackage{pifont}% http://ctan.org/pkg/pifont
\usepackage{amssymb}
\usepackage{afterpage}
\usepackage{capt-of}
\usepackage{bbm}
\usepackage{rotating}
\usepackage{lscape}
\usepackage{booktabs}
\usepackage[noindentafter]{titlesec}
\usepackage{hyperref}
\usepackage{pgfplots}
\pgfplotsset{width=8cm,compat=1.9}
\graphicspath{ {images/} }
\usepackage{amsmath}
\usepackage{braket} % needed for \Set
\usepackage{caption}
\usepackage{algorithm}
\usepackage{xcolor}
\usepackage{mathtools}
\DeclarePairedDelimiter\norm{\lVert}{\rVert}
\usepackage[noend]{algpseudocode}

\usepackage{mathtools,bm}

\def\BibTeX{{\rm B\kern-.05em{\sc i\kern-.025em b}\kern-.08em
    T\kern-.1667em\lower.7ex\hbox{E}\kern-.125emX}}
    
\makeatletter
\patchcmd{\@maketitle}
  {\addvspace{0.5\baselineskip}\egroup}
  {\addvspace{-1\baselineskip}\egroup}
  {}
  {}
\makeatother

\begin{document}
% \bstctlcite{IEEEexample:BSTcontrol}
\urlstyle{tt}

\title{PreGAN: Preemptive Migration Prediction Network for Proactive Fault-Tolerant Edge Computing}

\author{
        % \IEEEauthorblockN{Anonymous Authors}%
\IEEEauthorblockN{Shreshth Tuli\IEEEauthorrefmark{1}, Giuliano Casale\IEEEauthorrefmark{1}, Nicholas R. Jennings\IEEEauthorrefmark{1}\IEEEauthorrefmark{2}}
\IEEEauthorblockA{\IEEEauthorrefmark{1}{Imperial College London}}
\IEEEauthorblockA{\IEEEauthorrefmark{2}{Loughborough University}}
\IEEEauthorblockA{\{s.tuli20, g.casale\}@imperial.ac.uk, n.r.jennings@lboro.ac.uk}%
}

\maketitle
\thispagestyle{plain}
\pagestyle{plain}

\begin{abstract}
% Edge service reliability is an important issue that has emerged from the large-scale utilization of edge computing services to deploy and execute industrial applications. However, 
Building a fault-tolerant edge system that can quickly react to node overloads or failures is challenging due to the unreliability of edge devices and the strict service deadlines of modern applications. Moreover, unnecessary task migrations can stress the system network, giving rise to the need for a smart and parsimonious failure recovery scheme. Prior approaches often fail to adapt to highly volatile workloads or accurately detect and diagnose faults for optimal remediation. There is thus a need for a robust and proactive fault-tolerance mechanism to meet service level objectives. In this work, we propose PreGAN, a composite AI model using a Generative Adversarial Network (GAN) to predict preemptive migration decisions for proactive fault-tolerance in containerized edge deployments. PreGAN uses co-simulations in tandem with a GAN to learn a few-shot anomaly classifier and proactively predict migration decisions for reliable computing. Extensive experiments on a Raspberry-Pi based edge environment show that PreGAN can outperform state-of-the-art baseline methods in fault-detection, diagnosis and classification, thus achieving high quality of service. PreGAN accomplishes this by 5.1\% more accurate fault detection, higher diagnosis scores and 23.8\% lower overheads compared to the best method among the considered baselines.
%Specifically, PreGAN leads to reduced energy consumption, response time and SLO violations by up to 8, 5 and 12 percent, respectively.
\end{abstract}

%%%%%%%%%%%%%%%%%%%%%%%%%%%%%%%%%%%%%%%%%%%%%%%%%%%%%%%%%%%%%%%%%%%%%%%%%%%%%%%%
\section{Introduction}\label{sec:introduction}
\noindent
% Rise of IT systems and AI and increasing need of fault tolerance and fault detection/diagnosis/remediation
Edge computing is the processing of data close to the source where it is produced to optimize the service performance of such systems. This paradigm is closely integrated with the sensors and actuators in the Internet of Things (IoT) framework~\cite{gill2019transformative}.  With edge computing becoming ubiquitous, it is essential to ensure that the edge nodes themselves do not become a point-of-failure for the running applications, and robust countermeasures are in place to incorporate network or node overloads/failures. Modern application demands of low latency task execution and resource constraints of the edge devices further exacerbate the problem~\cite{tuli2020dynamic}. The increasing volumes of the data requiring immediate processing and the resource constraints at the edge are pushing the compute resources to their limits, giving rise to a high chance of resource contention and node downtimes~\cite{pcft, dastjerdi2016fog}. This leads to resource unavailability and Service Level Objective (SLO) violations that can lead to significant financial losses~\cite{nicoletti2013cloud}. Thus, it is critical to develop a fault-tolerance mechanism for edge computing to maintain low latency and high reliability.

% Challenges - diagnosis/root cause analysis in real time, volatile data, quick inference in time critical systems, low data availability in federated setups.
\textbf{Challenges.} The problem of developing a robust fault-tolerance framework is challenging. The first stage of solving this is to be able to proactively predict faults before they occur and diagnose the root-cause issues to be able to run appropriate remediation steps.  These steps should be carried out in near real-time for such systems to be effective~\cite{park2018lired}. Moreover, for such systems to be within the strict specifications of modern industrial demands, they need to be able to resolve diverse kinds of system or network related faults~\cite{malik2011adaptive}. This may entail establishing the type of fault at the time of its prediction for a more informed recovery decision. The volatile nature of the workloads and resources increases the difficulty in the prediction of faults and their types by several fold~\cite{ristov2020resilient}. Further, to avoid overlooking any fault that may cause significant adverse effects later and to avoid the overhead of false-positive predictions, such prediction models need to be extremely accurate. To do this, some existing methods leverage machine learning models due to their highly accuracy~\cite{eclb, cmodlb}. However, in such cases, the annotated log traces available to brokers is limited, leading to restricted size of the labelled fault classification data. This makes it hard to train supervised machine learning models for fault detection and classification.

% Prior works broadly and challenges - unstable, require lot of data, limited extraction/use of inter-correlations, limited multi-modal feature extraction, some dont use fault class knowledge, expensive. Cant use task replication due to limited compute resources at the edge, thus many works use preemptive migration.
\textbf{Existing solutions.} Over the past few years, several fault-tolerance approaches have been proposed to enhance the service reliability of edge or cloud platforms~\cite{pcft, dftm, kumari2018survey}. A popular method is to provide node redundancy and network contingencies to avoid the crippling downtime of an edge element. However, with the increasing number of edge devices, having redundancy for each node is not feasible, considering the energy and cost implications of such a deployment~\cite{zhou2010security}. Another way is to replicate the running instance of a task on a separate node~\cite{zheng2011component}. This is not ideal for resource-constrained edge devices as it makes them more susceptible to resource contention and faults~\cite{hong2019resource}.  Yet another mechanism is to periodically save the execution state of the running tasks by checkpointing their corresponding containers. Containers provide a virtualization layer that allows the running applications to be independent of the underlying hardware facilitating efficient task restoration in the event of node failures~\cite{khan2019edge}. When a node failure occurs, the system can transfer and resume the task on a different device, commonly referred to as ``preemptive migration''~\cite{engelmann2009proactive}. However, when predicting the faults and deciding the restoration node in advance, checkpointing the containers periodically could lead to excessive stress on the system and the network. Instead, we could checkpoint only those containers that need to be restored. This saves the excess overhead of periodically checkpointing all running tasks. In recent years, many different methods have been proposed to decide the appropriate migration decisions to enhance service reliability. These range from Particle Swarm Optimization (PSO)~\cite{pcft}, Integer Linear Programming~\cite{dftm, tian2018scheduling} and Bayesian Classification~\cite{eclb} to using Neural Network and clustering algorithms like k-means~\cite{cmodlb}. However, such methods are often not generalizable or struggle to adapt in volatile settings quickly, as discussed later in Section~\ref{sec:related_work}. Thus, we propose a novel fault-tolerance model and demonstrate its efficacy against state-of-the-art baselines~\cite{cmodlb, eclb}. 

% New insights - graph attentions, GRUs, prototype networks, self-attention, GANs. Thus we use GANs but that is insufficient -> prototype based classification, feature and time based feature extraction. 
\textbf{Background and new insights.}  To be able to predict an appropriate preemptive migration decision, it is crucial to accurately detect, diagnose and classify faults in an online fashion. One way to achieve this is to create a deep generator model that predicts such a decision. Generative Adversarial Networks (GANs) have been shown to be very successful in this as they can reduce prediction errors and provide us with a robust anomaly prediction framework~\cite{mad_gan}.  To successfully train a GAN, we need to efficiently model the temporal trends and the cross-correlations of performance and resource utilization metrics among different edge nodes. Recent developments in graph machine learning allow both of these to be done together efficiently~\cite{gat, gru}. To minimize the total computational complexity of each layer of the deep neural network and make decision prediction time-efficient, we can use self-attention operations~\cite{vaswani2017attention}. 
However, just having a GAN is insufficient to incorporate the fault types in the prediction framework. To be able to execute classification with limited data and allow quick adaptability, inspired from few-shot learning, we can extend prototype networks that generate an embedding vector for each class~\cite{snell2017prototypical}. Moreover, to test whether the preemptive migration decision would improve the Quality of Service (QoS), co-simulation techniques can be used~\cite{tuli2021cosco}. Due to the lack of integration interfaces, such advances have not been explored in the scope of edge reliability. This work leverages these concepts with necessary adaptations, to predict preemptive migration decisions for fault-tolerant edge computing.
% Each of these different insights cannot be directly used without intelligently integrating each component with necessary adaptations, which is done in this work. 

% In this paper, we propose ... - Need for each component (intuition) and final contribution, i.e. improvements.
\textbf{Our Contributions.} In this work, we propose \textbf{PreGAN}: \textbf{Pre}emptive Migration Prediction \textbf{GAN}. PreGAN uses graph attention and recurrent neural networks for feature extraction, a prototype network for classification and generates an input embedding for the GAN. The GAN-based prediction model enables robust training and time-efficient inference. We perform extensive empirical experiments on real-life edge testbed to compare and analyze PreGAN against the state-of-the-art methods. Our experiments show that PreGAN performs {best} in terms of QoS metrics, reducing the energy consumption, response time and SLO violations by up to 8, 5 and 12 percent, respectively. PreGAN achieves this by 5.1\% more accurate fault detection, 10.7\% lower task migrations  and 23.8\% lower overhead than the most accurate baseline.

The rest of the paper is organized as follows. Section~\ref{sec:related_work} overviews related work. Section~\ref{sec:method} outlines the PreGAN methodology for model training, preemptive migration prediction and execution. A performance evaluation of the proposed method is developed in Section~\ref{sec:experiments}. Finally, Section~\ref{sec:conclusion} concludes the paper and presents future directions.

% The rest of the paper is organized as follows ...

\section{Related Work}
\label{sec:related_work}

Fault-tolerance deals with developing systems that have an ability to withstand failures and faults in the workloads and efficiently manage resources to maintain optimum QoS. Most methods can be divided into two categories: reactive and proactive. The former take action after observing a system fault by typically checkpointing, replicating or resubmitting tasks affected by the fault~\cite{ataallah2015fault}. The latter aim to avoid expensive fault recovery by predicting failures in advance and taking appropriate steps for remediation by preemptive migration or fault-aware scheduling~\cite{ataallah2015fault}. As reactive schemes often lead to poor QoS in highly dynamic setups~\cite{pcft}, we focus on proactive methods in this work.

Most contemporary state-of-the-art techniques employ some form of specialized algorithms or machine learning models. Dynamic Fault Tolerant Migration (DFTM)~\cite{dftm} is a recent method that uses an integer linear programming (ILP) formulation to analyze workload traffic, select the tasks running on the hosts that should be migrated and the target hosts for restoration. Methods like Multi-stage Coflow Scheduling (MCS)~\cite{tian2018scheduling} also consider task co-dependencies to optimize QoS. Another ILP based method proposes a preference based fault management technique~\cite{pbfm} that tries to balance between the QoS improvement and the migration cost using a multi-objective formulation. Such methods do not scale for real-time operations, making them unsuitable for mission critical edge applications. SFS~\cite{mohammed2017failover} is a failover strategy that selects only those tasks that are about to violate their deadlines to reduce migration overheads. However, modeling the SLO deadlines does not consider other QoS parameters while optimizing the selection of the target hosts. This makes it often perform poorly in heterogeneous setups. 

The Proactive Coordinated Fault Tolerance (PCFT)~\cite{pcft} method uses Particle Swarm Optimization (PSO) to reduce the overall transmission overhead, network consumption and total execution time for a set of tasks. This method first predicts faults in the running host machines by anticipating resource deterioration and uses PSO to find target hosts for preemptive migration decision. This approach mainly focuses on reducing transmission overheads in distributed cloud setups, but often fails to improve the I/O performance of the compute nodes~\cite{dftm}. The Energy-efficient Checkpointing and Load Balancing (ECLB)~\cite{eclb} technique uses Bayesian methods and neural networks to classify host machines into three categories: overloaded, underloaded and normal execution. This classification is used to decide appropriate task migrations to reduce the number of overloaded hosts. However, this model only considers computational overloads and does not consider other fault types. CSAVM~\cite{satpathy2018crow} uses another evolutionary search scheme to take live migration decisions for the task queues. The method is used to optimize the power consumption of a compute setup by preventing unnecessary migrations. DDQP~\cite{wang2021ddqp} uses double deep Q-networks to place services on network nodes. However, such reinforcement learning schemes are known to be slow to adapt in volatile settings~\cite{tuli2021cosco}.

\setcounter{figure}{1}
\begin{figure*}[!t] 
    \centering  \setlength{\belowcaptionskip}{-12pt}
    \includegraphics[width=0.95\linewidth]{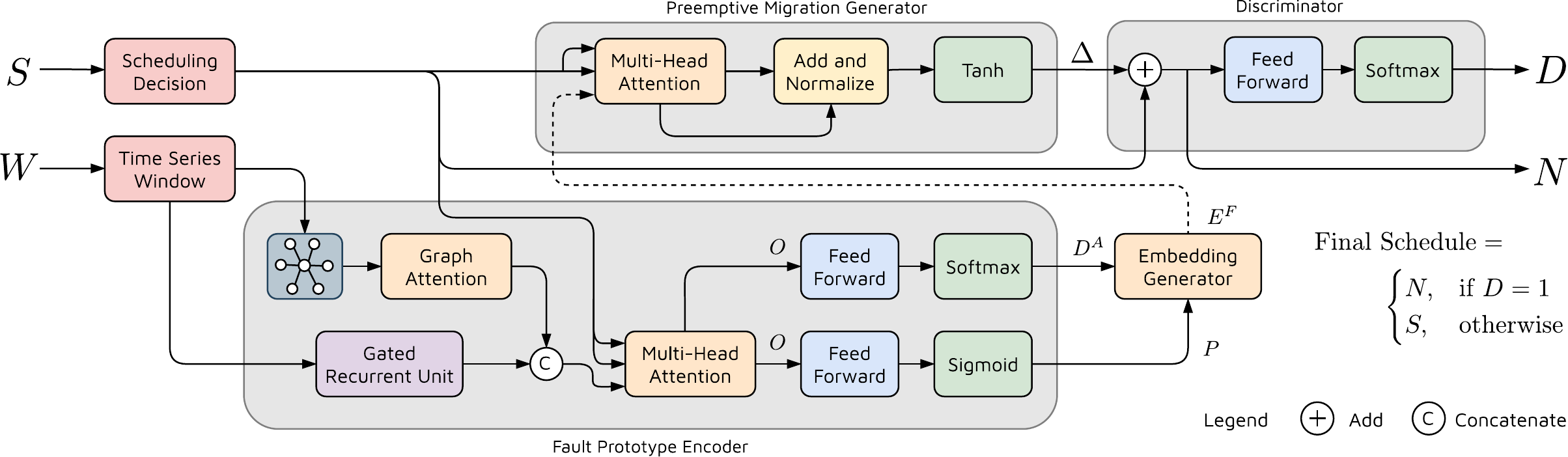}
    \caption{The PreGAN Model.}
    \label{fig:model}
\end{figure*}

\setcounter{figure}{0}
\begin{figure}
    \centering \setlength{\belowcaptionskip}{-10pt}
    \includegraphics[width=0.85\linewidth]{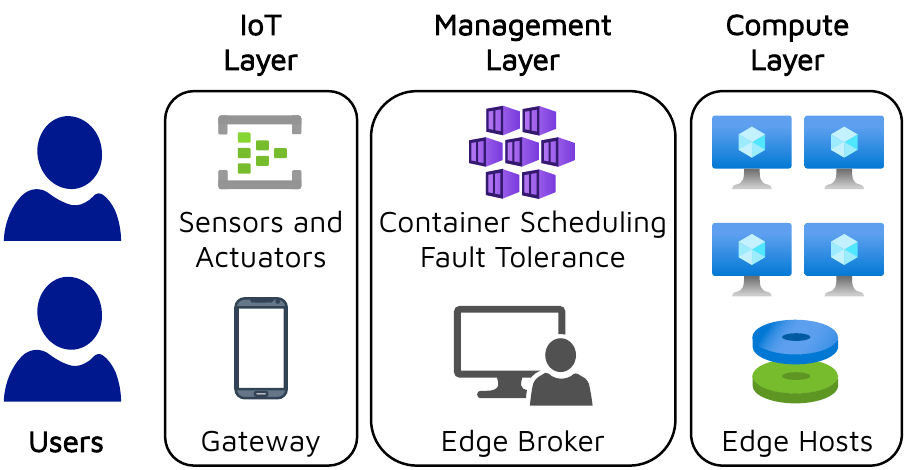}
    \caption{System Model.}
    \label{fig:system}
\end{figure}

\setcounter{figure}{2}

A recent work, CMODLB~\cite{cmodlb}, uses a clustering-based multiple objective dynamic load balancing technique to avoid resource contentions in cloud computing nodes. This method uses k-means to cluster nodes and identify overloaded hosts, PSO to select tasks and deep learning and fuzzy-logic optimization to select target hosts. %Tasks are selected from these hosts using the PSO approach and target hosts using deep learning and fuzzy-logic based optimization. 
However, slow PSO optimization leads to high recovery times that is often not helpful in highly dynamic systems~\cite{tuli2021cosco}. 
In our experiments, we compare PreGAN against the state-of-the-art baselines DFTM, ECLB, PCFT and CMODLB.

\section{Methodology}
\label{sec:method}

\subsection{System Model and Problem Formulation}
In this work, we target a standard heterogeneous edge computing environment where all nodes are in the same Local Area Network (LAN); see Figure~\ref{fig:system} for an overview. Tasks are container instances generated from the sensors and actuators in the IoT layer and communicated to the compute nodes via gateway devices. The edge broker takes all scheduling, management and preemptive migration decisions. 

We assume that there are a $m$ number of host machines in the fog resource layer and denote them as $H = \{h_1, \ldots, h_{m}\}$. We assume the paradigm of discrete time control, where we take scheduling and migration decisions at periodic intervals~\cite{tuli2019fogbus, basu2019learn, tuli2020dynamic}. We assume a scheduler is already present in the broker. The $t$-th interval is denoted as $I_t$ and the scheduling decision and host characteristics at this interval are denoted as $S_t$ and $x_t \in {\rm I\!R}^{m\times n}$ (each host has $n$ features). The time-series $\{x_0, \ldots, x_{t-1}\}$ is denoted as $\mathcal{T}_t$. We divide the problem of fault-tolerance into two sub-problems, defined as:

\textbf{Fault Prediction:} For an input multi-variate time-series $\{x_0, \ldots, x_{t-1}\}$, we need to predict fault labels for each host in $I_t$ as $y_t \in \{0, 1, \ldots, c\}^m$. Here, $y_{t,i}$ denotes the output for $h_i$; $y_{t,i} = 0$ means no fault and $y_{t,i} = j$ means fault of class $j$ among the user-specified $c$ classes $j \in \{1, \ldots, c\}$. 

\textbf{Preemptive Migration Decision:} For an input scheduling decision $S_t$ (set of task and host pairs) and a solution of the prediction problem $y_t$, we wish to predict a migration decision $\Delta_t$.
To model the dependence of a data point $x_t$ at a timestamp $t$, we consider the sliding window of length $k$, denoted as
\[W_t = \{x_{t-k}, \ldots, x_{t-1}\},\]
as is done in prior work~\cite{audibert2020usad}. Thus, in lieu of using just $x_t$, we use $W_t$ as it allows us to capture the local context. For model robustness, we normalize the input windows. For the sake of simplicity and without loss of generality, we drop the $t$ index whenever this is not ambiguous and use $x$, $S$, $W$, $\Delta$.

\subsection{PreGAN Model}

Figure~\ref{fig:model} provides an overview of the PreGAN model. The multi-variate time-series has now become the host characteristics $x$. We form a graph using the schedule $S$, such that there is an edge from $h_i$ to $h_j$ if there is a task migration from $h_i$ to $h_j$ in $S$. The $n$ characteristics of each host are then used to populate the feature vectors of the nodes in the graph. We then use a graph attention network (GAT) with Gated Recurrent Units (GRUs). GAT allows us to extract the multi-host feature correlations with the migration information encoded as graph edges. GRU allows us to extract temporal trends in time-series log data for prediction of faults in the next interval. This, in conjunction with the migration decision from the underlying scheduler, is used to detect the hosts with a high chance of faults and predict the class prototypes for each fault. The detection and classification results are combined to create an embedding for the generator network that predicts the preemptive migration decision to amend the input schedule. The decision is forwarded to a discriminator network to compare against the original scheduling decision.  

We now describe the complete pipeline in detail. The input time-series window $W$ is first converted to a tensor of size $k\times m\times n$. The scheduling decision for each task is converted to a one-hot decision vector of size $m$. For $p$ active tasks in the system, these one-hot vectors are stacked to form a matrix of size $p\times n$. We divide our model in two parts: (1) Fault Prototype Encoder (FPE) that aims to solve the first sub-problem and (2) Preemptive Migration GAN for the second sub-problem. 

\textbf{Fault Prototype Encoder.} To infer over all hosts in the graph, we create a new global node connected to each host node~\cite{tuli2021tango}. This gives the graph attention operation as
\begin{equation}
    O_1 = \mathrm{sigmoid} \bigg( \frac{1}{n} \sum_{i \in \{1, \ldots, m\}} \theta_{GAT} W_i \bigg),
\end{equation}
where $\theta_{\rm GAT}$ is the weight matrix for the GAT network. The time-series is also passed through a GRU, with $\bar{O}_2$ as the output of the previous interval
\begin{equation}
    O_2 = \mathrm{GRU}(W, \bar{O}_2).
\end{equation}
For any three input tensors $Q$, $K$ and $V$, we define Multi-Head Self Attention~\cite{vaswani2017attention} as passing it through $h$ (number of heads) feed-forward layers to get $Q_i$, $K_i$ and $V_i$ for $i \in \{1, \ldots, h\}$, and then applying attention as
\begin{align}
\begin{split}
    \mathrm{MultiHeadAtt}(Q, K, V) &= \mathrm{Concat}(H_1, \ldots, H_h)\\
     H_i &= \mathrm{Attention}(Q_i, K_i, V_i).
\end{split}
\end{align}
Here $\mathrm{Attention}(Q_i, K_i, V_i)$ is the scaled-dot product attention operation~\cite{vaswani2017attention}. We apply this on the GRU and GAT outputs (with task SLO requirements and host characteristics) to obtain
\begin{equation}
    O = \mathrm{MultiHeadAtt}(S, S, [O_1; O_2]).
\end{equation}
The late-fusion of the two outputs allows the downstream models to exploit them independently~\cite{tuli2021tango}.  The output $O$ is a collection of attention based encodings for each host. We pass this encoding through two decoders for each host $h_i$, 
\begin{align}
\begin{split}
    D^A_i &= \mathrm{softmax}(\mathrm{FeedForward}(O_i)),\\
    P_i &= \mathrm{sigmoid}(\mathrm{FeedForward}(O_i)),
\end{split}
\end{align}
where $D^A_i \in {\rm I\!R}^2$ denotes the fault prediction. The model predicts a fault in $h_i$ if $D^A_i[1] \geq D^A_i[0]$. $P_i \in {\rm I\!R}^E$ is a prototype embedding for each host corresponding to the fault class, irrespective of whether a fault is detected or not. This factored prediction for each host allows our model to be agnostic to the number of hosts in the system. We define the embedding for host $h_i$ as 
\begin{equation}
\label{eq:embedding_generator}
  E^F_i = \left\{
  \begin{array}{@{}ll@{}}
    P_i, & \text{if}\ D^A_i[1] \geq D^A_i[0]  \\
    \big[0\big]_{1\times E}, & \text{otherwise}.
  \end{array}\right.
\end{equation}
We stack all host embeddings to generate $E^F$. This auto-regressive style of using fault predictions allows us to generate a representation of the fault prediction that only consists of class prototypes for the hosts where a fault is detected.

\textbf{Preemptive Migration GAN.} The embedding output from the FPE is passed through the Generator
\begin{equation}
    \Delta\! =\! \mathrm{tanh}(\mathrm{LayerNorm}(S + \mathrm{MultiHeadAtt}(S, S, E^F))),
\end{equation}
where the $\Delta \in {\rm I\!R}^{p\times n}$ denotes the preemptive migration decision. This is added to the original scheduling decision to get $N = S + \Delta$. The final decision for each task $p$ is calculated as $\mathrm{argmax}(N_p)$. The output of the generator is passed to a discriminator decoder
\begin{equation}
    D = \mathrm{softmax}(\mathrm{FeedForward}(N)),
\end{equation}
where $D \in {\rm I\!R}^2$ denotes predicted likelihood scores for using $S$ and $N$ as schedules. The final scheduling decision of PreGAN is defined as
\begin{equation}
\label{eq:final_schedule}
  \mathrm{Final\ Schedule} = \left\{
  \begin{array}{@{}ll@{}}
    N, & \text{if}\ D[1] \geq D[0] \\
    S, & \text{otherwise}.
  \end{array}\right.
\end{equation}

Summarizing, the FPE detects faults in hosts and generates class prototype embeddings for each host. These embeddings are stacked ($E^F$) and sent to a generator network that outputs a preemptive migration vector ($\Delta$). The discriminator then provides a likelihood score to the new ($N$) and original ($S$) scheduling decisions. The final output is then the decision with the highest likelihood score.

\subsection{Offline FPE Training}

\begin{algorithm}[t]
    \begin{algorithmic}[1]
    \Require
    \Statex Fault Prototype Encoder $E$
    \Statex Dataset used for training $\{W_t, S_t, \hat{y}_t\}_{t=1}^T$
    \Statex Step size $\alpha$, Evolutionary hyperparameter $\epsilon$
    \Statex Iteration limit $L$
    \State Initialize weights in $E$. Set $l \gets 0$
    \State Randomly initialize class prototypes $\{P^C_0, \ldots, P^C_c\}$
    \State \textbf{do}
    \State \hspace{1em} \textbf{for}($t = 1 \text{ to } T$)
    \State \hspace{2em} $D^A, P \gets E(S_t, W_t)$ \label{line:inference}
    \State \hspace{2em} $L_1 = \sum_{i = 1}^m \big( \mathds{1}{(\hat{y}_{t,i} = 0)} \log (D^A[0])$
    \Statex \hspace{70pt} $+ \mathds{1}{(\hat{y}_{t,i} > 0)} \log (D^A[1]) \big)$ \label{line:l1}
    \State \hspace{2em} $L_2 = \sum_{i = 1}^m \mathds{1}{(\hat{y}_{t,i} > 0)} \big( \norm{P_i - P^C_{\hat{y}_{t,i}}}_2$ 
    \Statex \hspace{70pt} $-  ( \sum_{j \neq \hat{y}_{t,i}} \norm{P_i - P^C_j}_2 ) \big)$ \label{line:l2}
    \State \hspace{2em} \textbf{for}($i = 1 \text{ to } m$)
    \State \hspace{3em} \textbf{if}($\norm{P_i - P^C_{\hat{y}_{t,i}}}_2 = \min_j \norm{P_i - P^C_j}_2$)
    \State \hspace{4em} $P^C_{\hat{y}_{t,i}} \gets (1 - \alpha) \cdot P^C_{\hat{y}_{t,i}} + \alpha \cdot P_i$ \label{line:p_update}
    \State \hspace{2em} Update weights of $E$ using $L_1, L_2$ 
    \State \hspace{2em} $\alpha \gets (1 - \epsilon) \cdot \alpha$ \label{line:decay}
    \State \hspace{2em} $l \gets l + 1$
    \State \textbf{while} $l < L$
    \end{algorithmic}
\caption{The FPE training algorithm}
\label{alg:fpe_training} 
\end{algorithm}

% Existing datasets of fault labels, root causes and classes.
We now describe the training process for the FPE to detect and classify faults, summarized in Algorithm~\ref{alg:fpe_training}. To train the encoder, we first collect a dataset of input windows, scheduling decisions and fault class labels $\{W_t, S_t, \hat{y}_t\}_{t=1}^T$ by running the system without any preemptive migration. Here, $\hat{y}_{t,i}$ is the class label for $h_i$, with 0 indicating no fault. Now, the FPE encoder (denoted as $E$) generates the outputs $D^A$ and $P$ from the inputs $W_t$ and $S_t$ (line \ref{line:inference} in Alg.~\ref{alg:fpe_training}). We use the cross-entropy loss for fault detection (line~\ref{line:l1} in Alg.~\ref{alg:fpe_training}). Here, $\hat{y}_{t,i}$ is used to generate the ground-truth labels and the loss is summed over all hosts:
\begin{equation}
    L_1\! =\! \sum_{i = 1}^m \Big( \mathds{1}{(\hat{y}_{t,i}\! =\! 0)} \log (D^A[0])\! +\! \mathds{1}{(\hat{y}_{t,i}\! >\! 0)} \log (D^A[1]) \Big).
\end{equation}

For fault-classification, we use a triplet loss. We first define class prototype vectors for each fault class $\{P^C_0, \ldots, P^C_c\}$, that are randomly initialized from $[0, 1]^{E}$. Just like the means in a k-means clustering approach, these vectors denote the centroids of the embeddings of their respective classes~\cite{snell2017prototypical}. As is common in prototypical networks used in few-shot learning, the new embedding generated by the model belongs to the class which has the least Euclidean distance from the class prototypes. The triplet loss aims at reducing the distance of the output embedding from the true class prototype and increasing it from the false class ones (line~\ref{line:l2} in Alg.~\ref{alg:fpe_training}). Thus, for all those hosts with a true fault, the loss includes the L2-norm between the output embedding and the true class prototype and negative L2-norm between the output and false class prototypes:
\begin{equation}
    L_2\! =\! \sum_{i = 1}^m\! \mathds{1}{(\hat{y}_{t,i}\! >\! 0)} \Big( \norm{P_i - P^C_{\hat{y}_{t,i}}}_2  -  \sum_{j \neq \hat{y}_{t,i}}\! \norm{P_i - P^C_j}_2\! \Big).
\end{equation}

Another step in the training process is to update the class prototypes. If an output embedding belongs to the correct class, we can update the prototype of that class as
\begin{equation}
    P^C_{\hat{y}_{t,i}} \gets (1 - \alpha) \cdot P^C_{\hat{y}_{t,i}} + \alpha \cdot P_i,
\end{equation}
where $\alpha$ is the step-size. We exponentially decay our step-size to ensure that the model learns class prototypes quickly initially and converges to a stable prototype set (line \ref{line:decay} in Alg.~\ref{alg:fpe_training}). The triplet loss in conjunction with the prototype updates allow the prototypes to be distant from one another, making each prototype a characteristic identifier for that class.

\begin{figure}
    \centering  \setlength{\belowcaptionskip}{-15pt}
    \includegraphics[width=\linewidth]{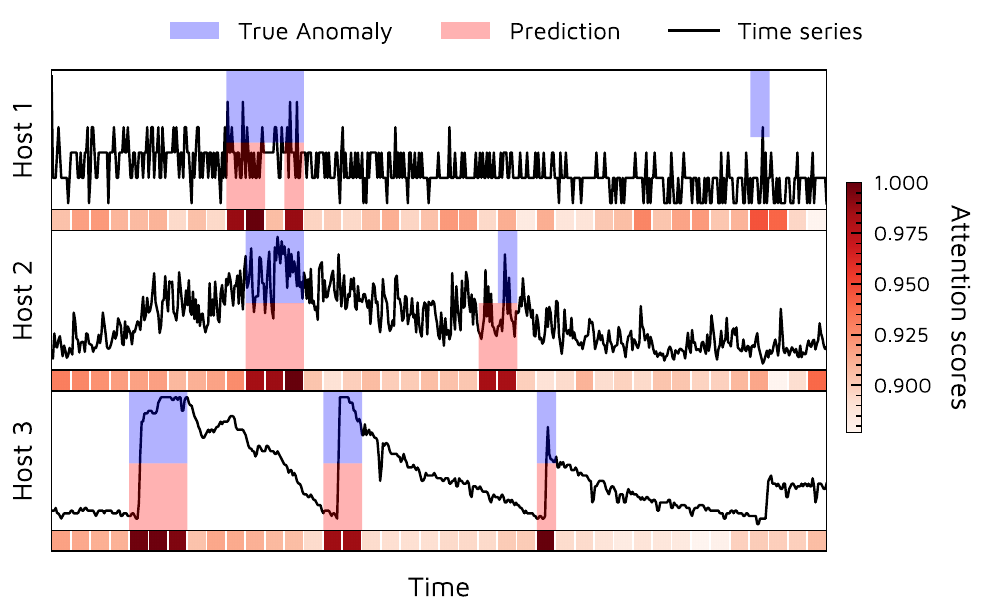}
    \caption{Visualization of Attention scores.}
    \label{fig:attention}
\end{figure}

\textbf{Visualization of Attention Scores.} Figure~\ref{fig:attention} visualizes the attention scores for the FPE encoder of the PreGAN model. The model is trained on a dataset collected from a real-setup (details in Section~\ref{sec:experiments}). We show the time-series, the average attention weights for each window (averaged over multiple heads and shown by the red heatmap) for the CPU utilization of the first 3 hosts in the system. It is apparent that the attention scores are highly correlated with the peaks in the data. This allows the model to specifically detect faults in each host individually. As shown in Figure~\ref{fig:attention}, the faults are detected in those hosts for which the attention scores are high.

\begin{algorithm}[t]
    \begin{algorithmic}[1]
    \Require
    \Statex Pretrained FPE Encoder $E$
    \Statex Generator and Discriminator networks $Gen$, $Disc$
    \Statex Co-Simulator $\mathrm{Sim}$
    \State Initialize weights in $Gen$, $Disc$.
    \State \textbf{for}($t = 1 \text{ to } T$)
    \State \hspace{1em} $D^A, P \gets E(S_t, W_t)$ 
    \State \hspace{1em} Get $E^F$ by \eqref{eq:embedding_generator} 
    \State \hspace{1em} $\Delta = Gen(S_t, E^F)$  \Comment{Generative preemptive migration}
    \State \hspace{1em} $N = S_t + \Delta$ \Comment{Form new scheduling decision}
    \State \hspace{1em} $D = Disc(S_t, N)$  \Comment{Compare the two decisions}
    \State \hspace{1em} Calculate $L_D$ using \eqref{eq:l3} \label{line:l3}
    \State \hspace{1em} Update weights of $Disc$ using $L_D$ 
    \State \hspace{1em} Calculate $L_G$ using \eqref{eq:l4} \label{line:l4}
    \State \hspace{1em} Update weights of $Gen$ using $L_G$ 
    \end{algorithmic}
\caption{The GAN training algorithm}
\label{alg:gan_training} 
\end{algorithm}

\subsection{Online GAN Training}
We now describe how we use a pre-trained FPE encoder and a co-simulation engine to train the GAN model, summarized in Algorithm~\ref{alg:gan_training}. Co-simulation is a technique in edge computing that allows a framework to run multiple event-based simulations with different parameters like scheduling decisions to optimize the QoS of the system~\cite{tuli2021cosco}. As event-based co-simulators can provide us QoS estimates quickly, they allow us to avoid running decisions on the physical setup, saving time and execution costs. Thus, unlike a traditional GAN model, we train our discriminator in a self-supervised manner using a co-simulator.

Our GAN training runs in an online fashion instead of using a pre-collected dataset.  For an input pair $(S, W)$, we run the PreGAN model to obtain $N$ and $S$. We run the co-simulator with the two scheduling decisions $S$ and $N$ and obtain the QoS scores. These scores may be calculated using a combination of metrics like energy consumption, response time and SLO violations~\cite{tuli2021hunter, basu2019learn}. We denote the co-simulated QoS scores by $\mathrm{Sim}(S)$ and $\mathrm{Sim}(N)$. To make the likelihood score output of the discriminator correspond to the co-simulated scores, we use the binary cross-entropy loss as:
\begin{align}
\label{eq:l3}
    &L_D\! =\! \frac{1}{2} \sum_{i = 1}^m \Big(  \mathds{1}{(\mathrm{Sim}(N)\! \geq\!  \mathrm{Sim}(S))} \Big(\!\log (D[1]) + \log (1 - D[0])\!\Big) \notag\\ &+\!  \mathds{1}{(\mathrm{Sim}(N)\! <\!  \mathrm{Sim}(S))} \Big(\! \log (D[0]) + \log (1 - D[1])\!\Big)\! \Big),
\end{align}
where the loss does not propagate to the generator (fixing $N$). This pushes the discriminator to give a higher likelihood score to the decision with a higher simulated QoS score. To train the generator, we use the adversarial loss
\begin{equation}
\label{eq:l4}
    L_G = \frac{1}{2} \sum_{i = 1}^m  \Big(\log (D[1]) + \log (1 - D[0])\Big),
\end{equation}
where the loss propagates to the generator with the discriminator weights kept fixed. This is equivalent to Eq.~\eqref{eq:l3} but with $\mathrm{Sim}(N)\! \geq\!\mathrm{Sim}(S)$, pushing the generator to output a migration decision $\Delta$ which gives a schedule $N$ better than the original schedule $S$. To do this, the generator gets fault class labels from the FPE encoder.

This style of training a discriminator model has two benefits: it allows us to (1) run our model inference without the co-simulation at test time, eliminating the computationally expensive generation of simulated traces, (2) train our preemptive migration generator by backpropagating the adversarial loss to the generator network.

\begin{algorithm}[t]
    \begin{algorithmic}[1]
    \Require
    \setlength{\textfloatsep}{0pt}
    \Statex Pretrained models $E$, $Gen$, $Disc$
    \State \textbf{for}($t = 1 \text{ to } T$)
    \State \hspace{1em} $D^A, P \gets E(S_t, W_t)$ 
    \State \hspace{1em} Get $E^F$ by \eqref{eq:embedding_generator}
    \State \hspace{1em} $\Delta = Gen(S_t, E^F)$  \Comment{Generative preemptive migration}
    \State \hspace{1em} $N = S_t + \Delta$ \Comment{Form new scheduling decision}
    \State \hspace{1em} $D = Disc(S_t, N)$  \Comment{Compare the two decisions}
    \State \hspace{1em} Execute final schedule obtained from \eqref{eq:final_schedule}
    \State \hspace{1em} \textbf{if}($D[1] \geq D[0]$) \label{line:if}
    \State \hspace{2em} \textbf{return} $N$ \Comment{Return new decision if higher score}
    \State \hspace{1em} \textbf{return} $S$ \label{line:s}
    \end{algorithmic} 
\caption{The PreGAN testing algorithm}
\label{alg:testing}
\end{algorithm}

\subsection{Online Inference}
We now describe the inference procedure using the trained PreGAN model (summarized in Algorithm~\ref{alg:testing}). For any input time-series window and scheduling decision $W, S$, the PreGAN model outputs $D, N$. Based on the likelihood scores of the discriminator $D$, PreGAN predicts if the preemptive migration decision would improve QoS or not. If it does, \textit{i.e.}, $D[1] \geq D[0]$, $N$ is executed, otherwise the original decision $S$ is executed (lines~\ref{line:if}-\ref{line:s} in Alg.~\ref{alg:testing}).

Overall, the FPE in PreGAN allows us proactively predict faults and guide the GAN model to generate a preemptive migration decision over the schedule generated by a policy oblivious to future system faults. The discriminator of the GAN then performs a cost-benefit analysis over the original scheduling decision to avoid excessive migration overheads. This allows PreGAN to optimize scheduling decisions and cut down execution costs.

\section{Experiments}
\label{sec:experiments}
We compare the PreGAN method with the state-of-the-art baselines: DFTM~\cite{dftm}, ECLB~\cite{eclb}, PCFT~\cite{pcft} and CMODLB~\cite{cmodlb} (more details in Section~\ref{sec:related_work}). Other heuristic based approaches have been omitted for brevity as these baselines outperform those in all the metrics we consider in our experiments. We use hyperparameters of the baseline models as presented in their respective papers. We train all deep learning models using the PyTorch-1.8.0~\cite{paszke2019pytorch} library\footnote{All model training and experiments were performed on a system with configuration: Intel i7-10700K CPU, 64GB RAM, Nvidia RTX 3080 and Windows 10 Pro OS. This was also the broker node in our setup.}. 

\subsection{Evaluation Setup}
\label{sec:setup}
Our evaluation testbed is a cluster with 16 Raspberry Pi 4B nodes.% (see Figure~\ref{fig:rpi}). 
Our cluster contains 8 nodes with 4-GB RAM and another 8 but with 8-GB RAM. %The gateway devices are considered to be within the same LAN. 
The power consumption models are taken from the commonly-used SPEC benchmarks repository~\cite{spec}. We run all experiments for 100 scheduling intervals, with each interval being 300 seconds long, giving a total experiment time of 8 hours 20 minutes. We average over 5 runs and use diverse workload types to ensure statistical significance.

For our workloads, we consider the \textit{DeFog} applications: Yolo, PocketSphinx and Aeneas~\cite{mcchesney2019defog}. DeFog is a fog computing benchmark suite that consists of various real-world application instances. The three specific applications used in our experiments were considered due to their heterogeneous resource requirements and volatile characteristics. At the beginning of every scheduling interval, we create $Poisson(\lambda)$ new workloads, sampled uniformly from the three applications. Poisson distribution is a natural choice for a bag-of-tasks workload model, common in edge environments~\cite{mao2016dynamic, basu2019learn, tuli2020dynamic}.

% \begin{figure}[t]
%     \centering \setlength{\belowcaptionskip}{-12pt}
%     \includegraphics[width=0.45\linewidth]{cluster_short.jpg}
%     \caption{Raspberry Pi Cluster as evaluation platform.}
%     \label{fig:rpi}
% \end{figure}
%

\subsection{Evaluation Metrics}
% detection - accuracy, p, r, f1. diagnosis - hr, ndcg. overhead ratio. improvement ratio (migration figure).
% qos - energy, response time, SLO, migration time, CPU/RAM utilization.
For fault detection, we consider detection accuracy, precision, recall and F1 scores. For a test datapoint $\{W_t, S_t, \hat{y}_t\}$ with the PreGAN outputs $D$ and $N$, the predicted and ground-truth labels are obtained as \[\bigwedge_{i = 1}^m \mathds{1}{(D[1] > D[0])} \text{ and } \bigwedge_{i = 1}^m \mathds{1}{(\hat{y}_{t,i} > 0 )}\]
respectively. This ground-truth label is 1 when any of the edge hosts has a fault and 0 otherwise.

For fault diagnosis, we consider two commonly used metrics~\cite{mtad_gat}: (1) $\mathrm{HitRate@100\%}$ is the measure of how many ground truth dimensions have been included in the top candidates predicted by the model~\cite{omnianomaly}, (2) Normalized Discounted Cumulative Gain ($\mathrm{NDCG@100\%}$)~\cite{jarvelin2002cumulated}. For fault classification, we consider the classification accuracy. For test time, we also consider an "improvement ratio", which for an execution of T intervals is calculated as:
\begin{equation}
    \text{Improvement Ratio} = \frac{1}{T} \sum_{t = 1}^T \mathds{1}{( D[1] > D[0] )}.
\end{equation}
\begin{figure}
    \centering \setlength{\belowcaptionskip}{-10pt}
    \includegraphics[width=0.8\linewidth]{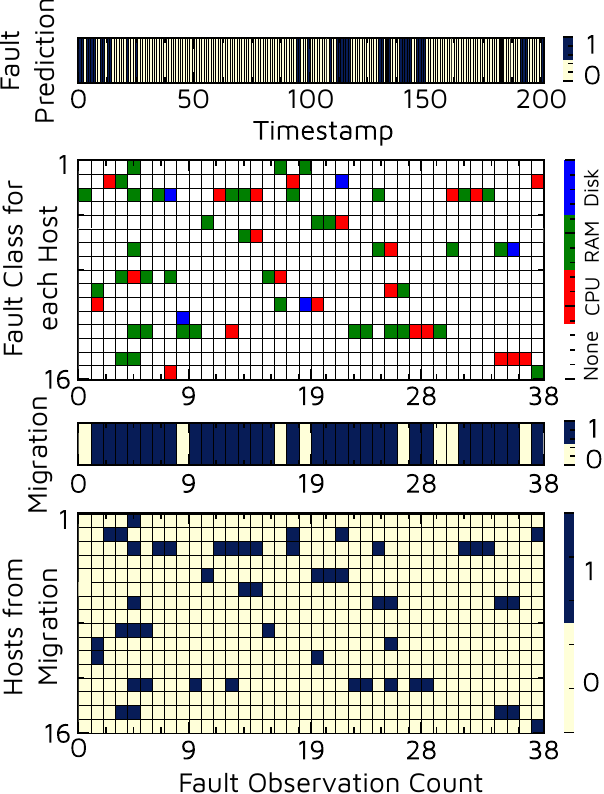}
    \caption{Visualization of fault prediction, classification and migration decisions with execution intervals.}
    \label{fig:migrations}
\end{figure}
This metric denotes the ratio of the times the model is able to predict a better scheduling decision than the original. This gives us an indication of the how well the preemptive migration prediction performs compared to the original scheduling decision. Together with the detection accuracy this gives us a complete picture on how well the model performs compared to the case where there is no preemptive migration. To explain these metrics with a simple visualization example on 200 intervals, consider Figure~\ref{fig:migrations}. The top heatmap denotes the intervals at which the model predicted a fault (fault denoted with black and yellow otherwise), specifically 38 intervals out of 200. The second heatmap shows the fault class prediction for each host (if any). The third heatmap shows the intervals in which the likelihood score of $N$ was higher than $S$ (30 out of 38) and the fourth heatmap shows the hosts from which there was a task migrated.  Here, the improvement ratio is $30 / 38 = 0.7895$. Also, we observe a strong correlation between the fault class predictions and the host selected for task migration, thanks to the factored prediction in PreGAN.

We also consider standard QoS metrics including energy consumption, response time, fraction of SLO violations, migration count and resource utilization.

\begin{figure}[t]
    \centering \setlength{\belowcaptionskip}{-15pt}
    \includegraphics[width=0.85\linewidth]{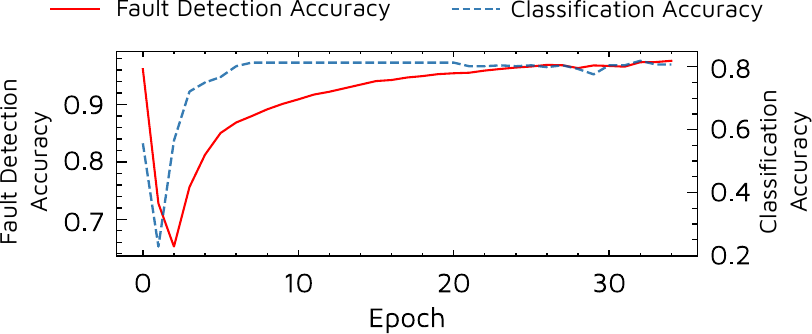}
    \caption{Training the Fault Prototype Encoder.}
    \label{fig:accuracy}
\end{figure}
\begin{figure}
    \centering \setlength{\belowcaptionskip}{-15pt}
    \includegraphics[width=0.75\linewidth]{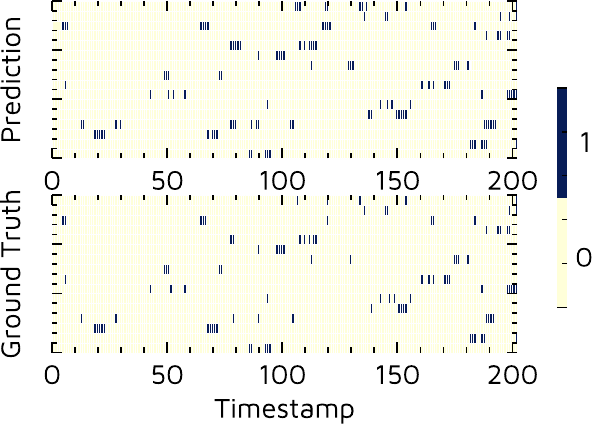}
    \caption{Fault Prediction with Ground Truth labels.}
    \label{fig:scores}
\end{figure}

\begin{figure}[t]
    \centering \setlength{\belowcaptionskip}{-12pt}
    \subfigure[t-SNE plot of prototypes]{
    \raisebox{7pt}{\includegraphics[width=0.47\linewidth]{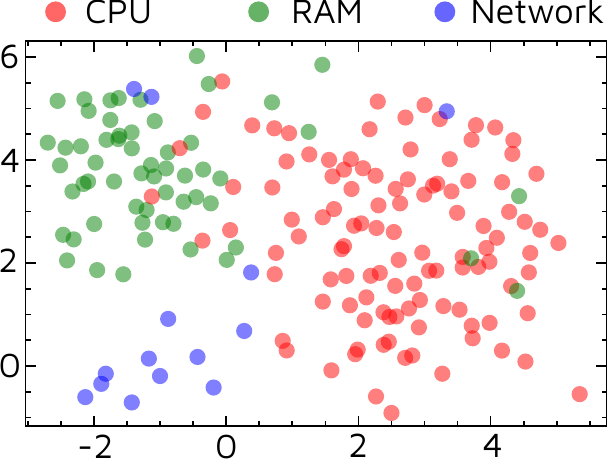}}
    \label{fig:tsne}
    }%
    \subfigure[Confusion Matrix]{
    \includegraphics[width=0.49\linewidth]{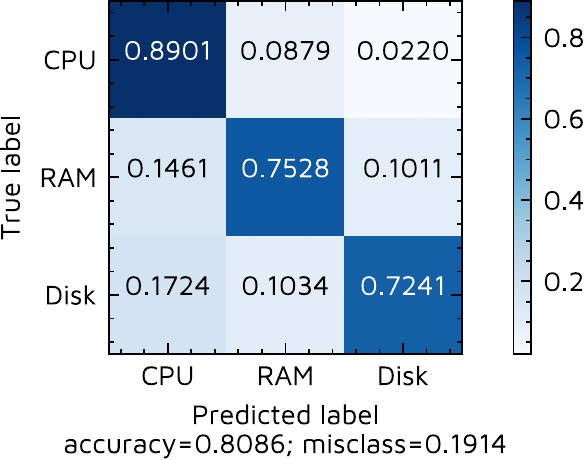}
    \label{fig:confusion}
    }
    \caption{Fault Classification on the test set.}
\end{figure}
\begin{figure}[h!]
    \centering \setlength{\belowcaptionskip}{-15pt}
    \includegraphics[width=0.8\linewidth]{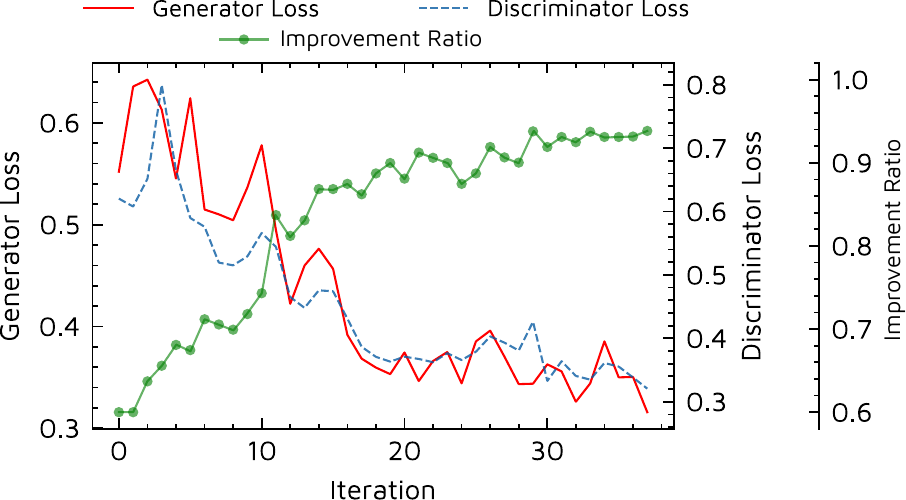}
    \caption{Training of the GAN network.}
    \label{fig:gan_training}
\end{figure}

\subsection{Implementation and Training}
% c classes, n features. FPE training (dataset, hyperparameters - alpha, epsilon). Visualization of detection and classification (tsne). 
% GAN training (qos score calculation). 
To conduct our tests, we use the COSCO framework that supports container orchestration in distributed edge clusters~\cite{tuli2021cosco}. COSCO is at present the only framework that allows the generation of QoS scores using co-simulated traces. For a fair comparison, we use a common underlying scheduling policy, GOBI~\cite{tuli2021cosco}, that generates scheduling decisions ($S_t$) by optimizing QoS scores using a deep neural network based surrogate model. %GOBI is a recent scheduling approach that is shown to be applicable to diverse use-cases and outperforms other reinforcement learning and heuristics based methods~\cite{tuli2021cosco}.

\begin{table*}[!t]
    \centering
    \caption{Performance scores of PreGAN and the baseline methods with standard deviation. The best scores are shown in bold.}
    \resizebox{\textwidth}{!}{
    \begin{tabular}{@{}lcccccccc@{}}
    \toprule 
    \multirow{2}{*}{Method} & \multicolumn{4}{c}{Detection} & \multicolumn{2}{c}{Diagnosis} & \multirow{2}{*}{\begin{tabular}{@{}c@{}}Overhead \\ Ratio\end{tabular}} & \multirow{2}{*}{\begin{tabular}{@{}c@{}}Improvement \\ Ratio\end{tabular}}\tabularnewline
    \cmidrule{2-7}
     & Accuracy & Precision & Recall & F1 Score & HR@100 & NDCG@100 & & \tabularnewline
     \midrule
    DFTM & 0.8731 \scriptsize{$\pm$0.0234} & 0.7713 \scriptsize{$\pm$0.0823} & 0.8427 \scriptsize{$\pm$0.0199} & 0.8054 \scriptsize{$\pm$0.0872} & 0.5129 \scriptsize{$\pm$0.0212} & 0.4673 \scriptsize{$\pm$0.0019} & \textbf{0.0413 \scriptsize{$\pm$0.0021}} & 0.3783 \scriptsize{$\pm$0.1001}\tabularnewline
    ECLB & 0.9413 \scriptsize{$\pm$0.0172} & 0.7812 \scriptsize{$\pm$0.0711} & 0.8918 \scriptsize{$\pm$0.0203} & 0.8329 \scriptsize{$\pm$0.0901} & 0.4913 \scriptsize{$\pm$0.0010} & 0.5239 \scriptsize{$\pm$0.0024} & 0.1028 \scriptsize{$\pm$0.0009} & 0.5912 \scriptsize{$\pm$0.0341}\tabularnewline
    PCFT & 0.8913 \scriptsize{$\pm$0.0108} & 0.8029 \scriptsize{$\pm$0.0692} & \textbf{0.9018 \scriptsize{$\pm$0.0165}} & 0.8495 \scriptsize{$\pm$0.0312} & 0.5982 \scriptsize{$\pm$0.0094} & 0.5671 \scriptsize{$\pm$0.0020} & 0.0913 \scriptsize{$\pm$0.0014} & 0.6824 \scriptsize{$\pm$0.0473}\tabularnewline
    CMODLB & 0.9128 \scriptsize{$\pm$0.0112} & 0.8158 \scriptsize{$\pm$0.0343} & 0.9013 \scriptsize{$\pm$0.0091} & 0.8605 \scriptsize{$\pm$0.0284} & \textbf{0.6309 \scriptsize{$\pm$0.0025}} & 0.5432 \scriptsize{$\pm$0.0031} & 0.2123 \scriptsize{$\pm$0.0003} & 0.7283 \scriptsize{$\pm$0.0065}\tabularnewline
    \textbf{PreGAN} & \textbf{0.9635 \scriptsize{$\pm$0.00921}} & \textbf{0.8723 \scriptsize{$\pm$0.0221}} & \textbf{0.9018 \scriptsize{$\pm$0.0121}} & \textbf{0.8868 \scriptsize{$\pm$0.0629}} & 0.6232 \scriptsize{$\pm$0.0069} & \textbf{0.5898 \scriptsize{$\pm$0.0080}} & 0.1617 \scriptsize{$\pm$0.0017} & \textbf{0.7605 \scriptsize{$\pm$0.0060}}\tabularnewline
    \bottomrule 
    \end{tabular}} \vspace{-1.7em}
    \label{tab:results}
\end{table*}
\begin{figure*}[!t]
    \centering
    \setlength{\belowcaptionskip}{-8pt}
    \subfigure[Energy Consumption]{
    \includegraphics[width=.23\textwidth]{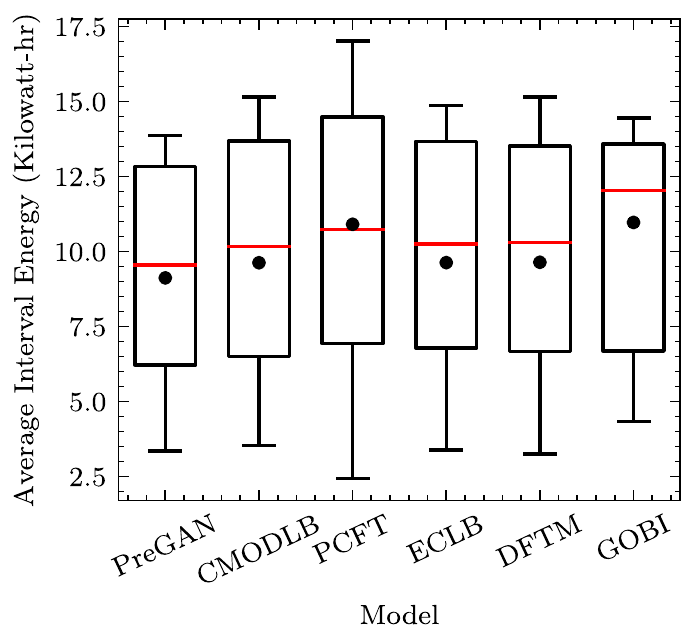}
    \label{fig:energy}
    }
    \subfigure[Response Time]{
    \includegraphics[width=.23\textwidth]{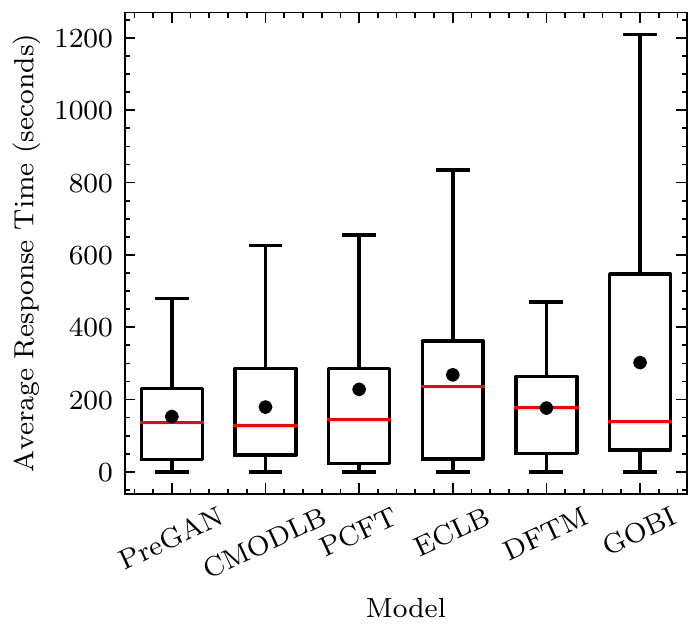}
    \label{fig:response}
    }
    \subfigure[CPU Utilization]{
    \includegraphics[width=.23\textwidth]{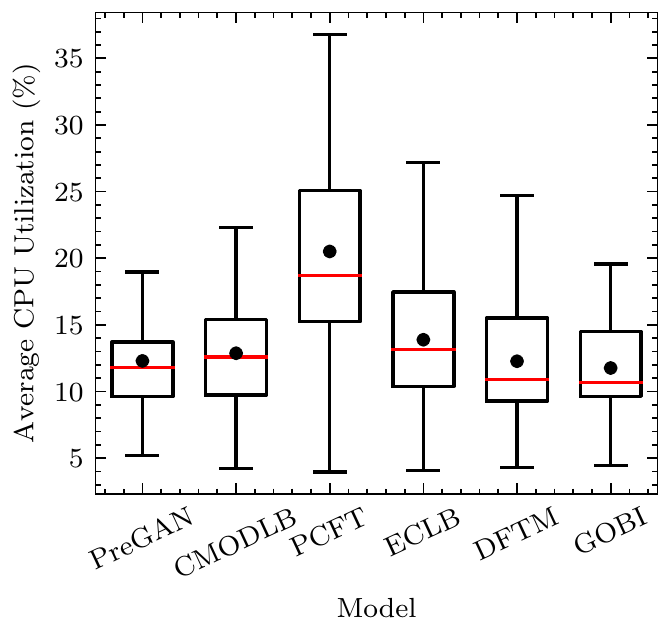}
    \label{fig:cpu}
    }
    \subfigure[RAM Utilization]{
    \includegraphics[width=.23\textwidth]{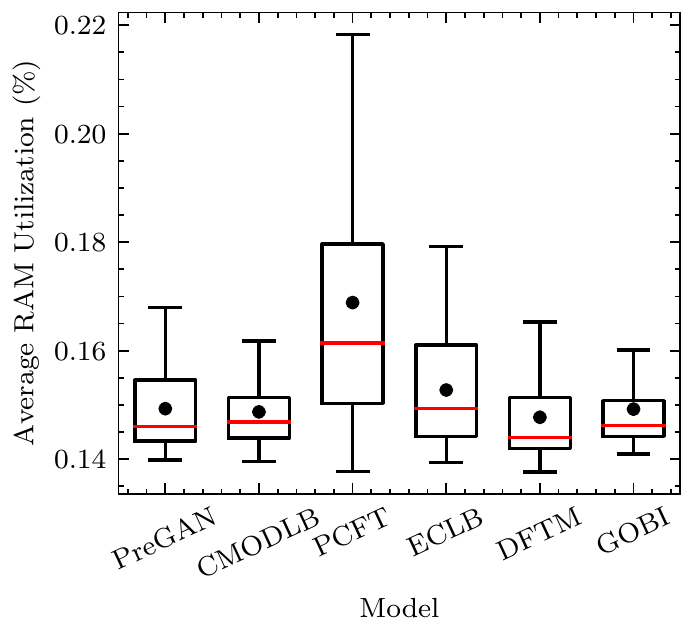}
    \label{fig:ram}
    }\\
    \subfigure[Migration Time vs Interval]{
    \raisebox{11pt}{\includegraphics[width=.23\textwidth]{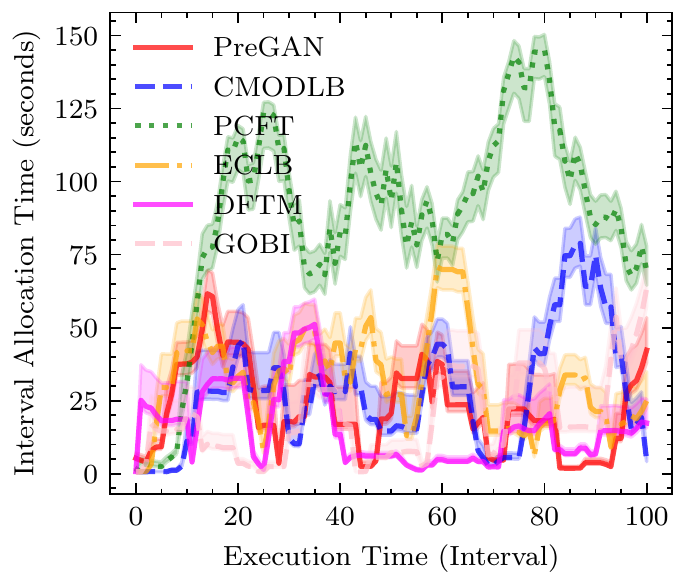}}
    \label{fig:alloc_time}
    }
    \subfigure[Migration Counts]{
    \includegraphics[width=.23\textwidth]{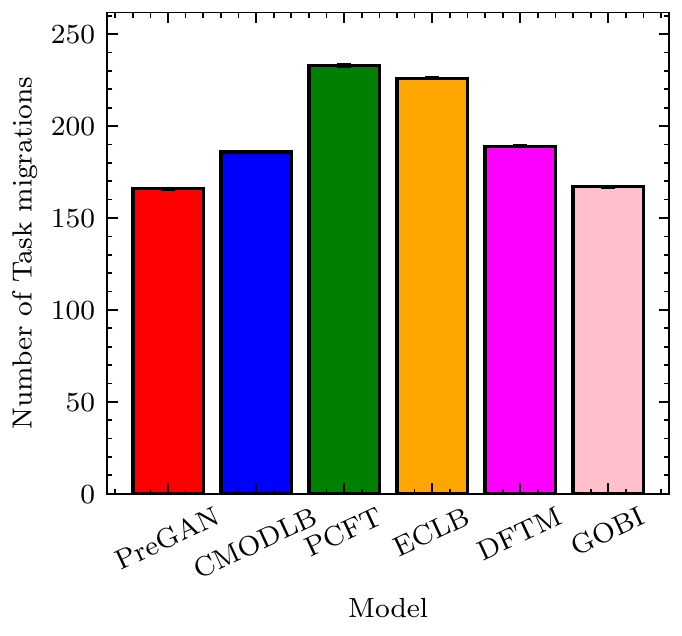}
    \label{fig:migrations}
    }
    \subfigure[Fraction of SLO Violations]{
    \includegraphics[width=.23\textwidth]{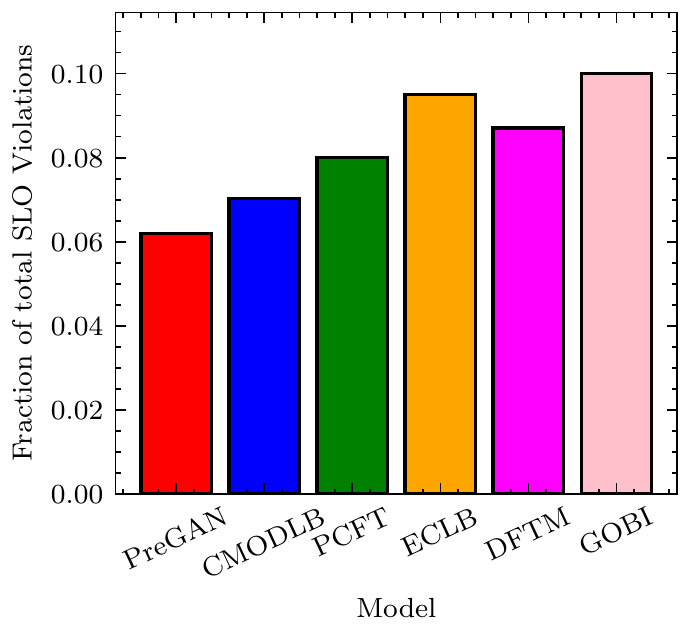}
    \label{fig:sla}
    }
    \subfigure[SLO Violations (per application)]{
    \includegraphics[width=.23\textwidth]{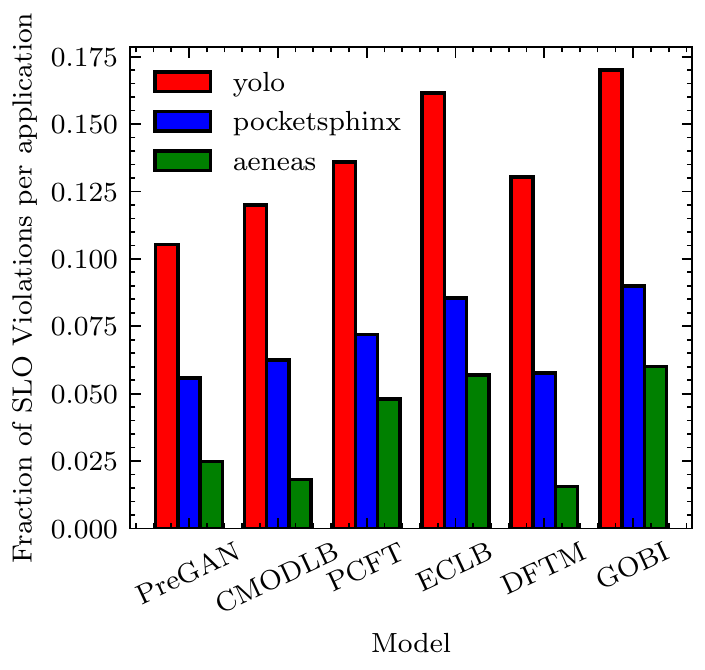}
    \label{fig:sla_pa}
    }
    \caption{Comparison of QoS parameters of PreGAN against baselines.} \vspace{-1em}
    \label{fig:results}
\end{figure*}

To generate the ground-truth fault labels and classes, we use the Anomaly Detection Engine for Linux Logs (ADE) tool~\cite{ade}. The output anomaly classes for the tool are: CPU over-utilization (CPUO), Abnormal disk utilization (ADU), Memory leak (MEL), Abnormal memory allocation (AMA) and Network overload (NOL). CPUO occurs when the CPU utilization exceed 90\%. When the disk controller of a system throttles read/write operations, an ADU fault occurs. MEL occurs when there is incremental allocation of 1 MB memory every 3 seconds. When RAM I/O utilization exceed 90\%, it causes an AMA fault. NOL fault occurs when network buffer overloads and temporary disk space needs to be allocated for file transfers. All faults events are raised when the above conditions hold for at least 60 seconds. %As we consider an edge cluster in a LAN, and thus with high network capacity, we ignore the network overload labels. 
For simplicity, the CPUO anomaly is classified as a CPU fault, NOL/ADU are clubbed together as a Network fault (as network buffer overloads almost always result in disk faults) and MEL/AMA are clubbed together as a RAM fault (due to high correlation between these two). Thus, we consider $3$ fault types/classes.

To train the FPE and GAN models, we use the AdamW optimizer~\cite{saleh2019dynamic}. We train the FPE encoder and GAN using a traces of lengths 1000 and 1200 scheduling intervals respectively. These numbers were obtained using the early-stopping convergence criterion. The QoS score obtained from the co-simulations was a convex combination of the normalized energy consumption and SLO violations~\cite{tuli2021cosco}.  The hyperparameter values used in our experiments for PreGAN were obtained using grid search. We use the prototype embedding size of $8$, windows size of $5$, initial value of $\alpha$ as $0.9$ and decay rate $\epsilon$ as $0.05$. 

Figure~\ref{fig:accuracy} shows how the FPE encoder converges in offline training using $80\%$ of the dataset collected from traces of 1000 intervals (rest as test set). As the model is trained, its fault detection and classification accuracies improve. Figure~\ref{fig:scores} shows the predicted and ground-truth fault labels on the test set, with hosts on the y-axis and intervals on the x-axis. Figure~\ref{fig:tsne} shows a t-SNE plot of the class prototypes predicted on the test set. Clearly, the prototype embeddings of the same class are clustered together, demonstrating that the model is able to distinguish among the different fault types. Classification confusion matrix on the test set can be seen in Figure~\ref{fig:confusion}. The prototype embeddings generation allows the model to generalize and accurately classify even previously unseen time-series inputs into one of the three classes. After training the FPE, we train the GAN. Figure~\ref{fig:gan_training} shows the discriminator ($L_D$) and generator ($L_G$) loss values with the the Improvement Ratio for the first 40 iterations, \textit{i.e.}, the intervals with faults in the GAN training dataset.

\begin{figure*}[!t]
    \centering \setlength{\belowcaptionskip}{-12pt}
    \includegraphics[width=.45\textwidth]{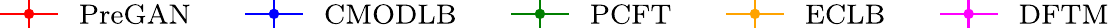}\\
    \subfigure[Energy Consumption]{
    \includegraphics[width=.23\textwidth]{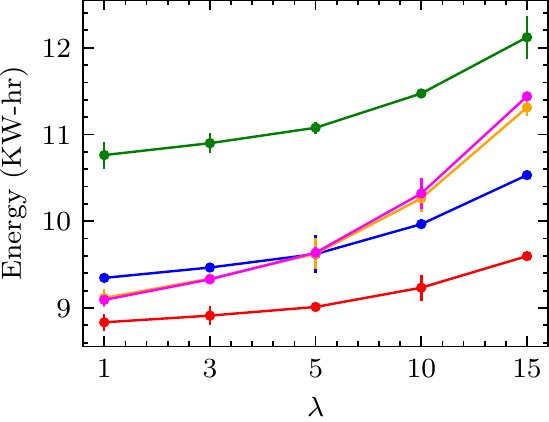}
    \label{fig:s_energy}
    }
    \subfigure[Fraction of SLO Violations]{
    \includegraphics[width=.23\textwidth]{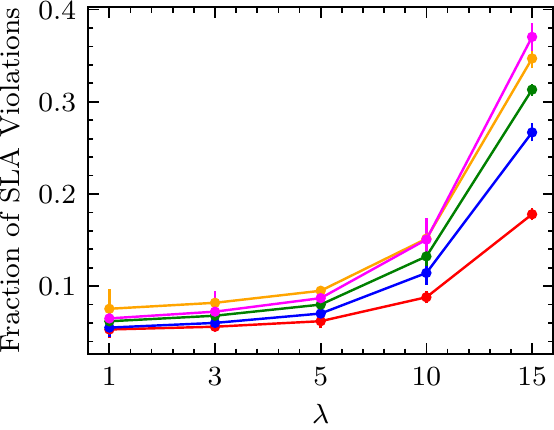}
    \label{fig:s_sla}
    }
    \subfigure[F1 Score]{
    \includegraphics[width=.23\textwidth]{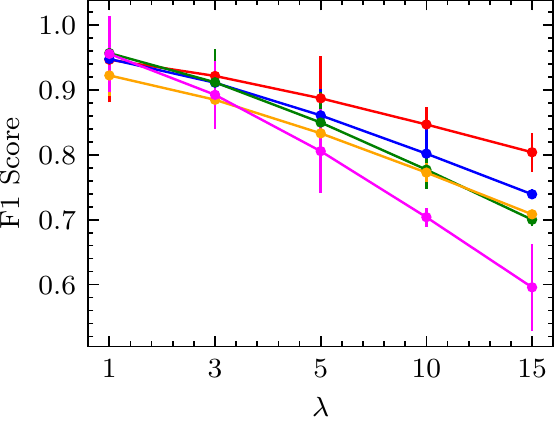}
    \label{fig:s_f1}
    }
    \subfigure[Improvement Ratio]{
    \includegraphics[width=.23\textwidth]{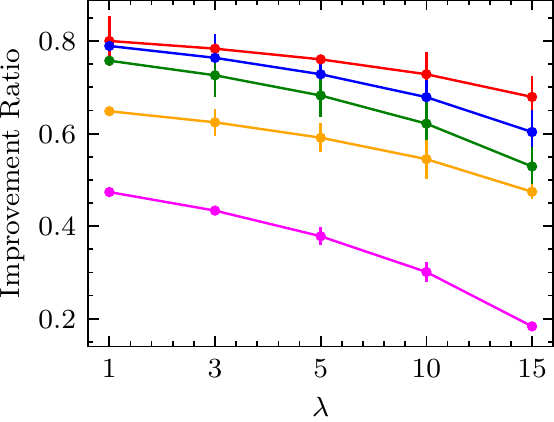}
    \label{fig:s_impr}
    }
    \caption{Sensitivity Analysis for all models with $\lambda$ (parameter of the discrete Poisson distribution).} \vspace{-0.7em}
    \label{fig:sens}
\end{figure*}

\vspace{-1pt}
\subsection{Results}
We now describe the results corresponding to 100 interval runs on the testbed using the DeFog workloads. The ground-truth labels are obtained offline, after the complete execution to compare the detection and diagnosis performance of the models. Table~\ref{tab:results} shows the detection and diagnosis metrics with the improvement ratios. It also shows the overhead ratio in terms of the time it takes to run the preemptive migration model relative to the scheduling model. The overhead ratio is highest for CMODLB due to the several sequential steps in its model like k-means clustering, ANN inference, and PSO. On average, the accuracy, precision, recall and the F1 score of the PreGAN model are the highest. This is due to the late-fusion of the embeddings obtained by the feature (GAT) and temporal trend (GRU) extraction. This allows the PreGAN model to not only exploit the correlations across host devices, but also sudden deviations from the expected resource utilization characteristics across time. Models like DFTM and ECLB use a basic sinusoidal thermal trend for all models and do not consider host heterogeneity. CMODLB uses k-means clustering to identify a common utilization threshold, irrespective of the running workloads or the host capacities. This gives higher fault detection scores for the PreGAN model. For fault-diagnosis, the CMODLB method has the highest HitRate (0.6309) with the PreGAN being very close (0.6232). The NDCG score of the PreGAN model is the highest. This is due to the factored fault prediction in the PreGAN model. 
%Finally, the improvement ratio of the PreGAN model is the highest, improving from the CMODLB model by 4.42\%. 
Finally, the improvement ratio of the PreGAN model is the highest, \textit{i.e.}, 0.7605 improving from 0.7283 of the CMODLB model by 4.42\%.

Figure~\ref{fig:results} compares the QoS scores of all models, including the vanilla GOBI approach without any preemptive migrations. The PreGAN model has the lowest average energy consumption of 9.0121 KW-hr, with DFTM being next at 9.4255 KW-hr. This is due to the relatively lower average CPU and RAM utilizations (Figures~\ref{fig:cpu} and \ref{fig:ram}). Figure~\ref{fig:response} shows the average response times with response time being defined as the time between the creation of a task from an IoT sensor and the gateway receiving the response. Among the baselines, the DFTM approach has the lowest response time as it uses minimum-migration-time heuristic~\cite{beloglazov2012optimal} for task migrations (Fig.~\ref{fig:alloc_time}). PreGAN avoids unnecessary migrations to prevent avoidable use of network resources, improving overall system reliability (Fig.~\ref{fig:migrations}).  Even with higher response times, CMODLB and PreGAN achieve low SLO violation rates as their migration decisions are SLO aware. Figure~\ref{fig:sla_pa} shows the SLO violation rates for each application. For all models, Yolo has the highest violation rate due to its compute heavy utilization characteristics. The migration decisions in PreGAN use the fault class labels to decide the target hosts in the migration decisions. The fine-grained classification in PreGAN allows it to migrate CPU intensive tasks from a compute constrained node to a node with low CPU utilization, even if the target node is running at capacity on RAM and Disk. Compared to the binary classification in CMODLB, this gives PreGAN more choices for the target host, allowing lower overheads in migration and more balanced resource utilization.

\subsection{Sensitivity Analysis}
\label{sec:analyses}

Figure~\ref{fig:sens} shows the variation of the energy consumption, SLO violations, F1 score and improvement ratio with the $\lambda$ parameter in our Poisson distribution used to model the workloads. We vary $\lambda$ from 1 to 15 ($\lambda = 15$ constantly gives $>90\%$ CPU utilization for all hosts). Under a higher $\lambda$ more tasks are produced, making the fault prediction harder. This is apparent from the drop in the F1 scores, leading to higher SLO violations. Even the energy consumption increases due to the increase in the average CPU utilization of the system. Overall, PreGAN shows the least relative drop in F1 scores and improvement ratio as we increase $\lambda$ giving the least SLO violations even in workload heavy executions.

\section{Conclusions}
\label{sec:conclusion}
We have presented a preemptive migration prediction model (PreGAN) that can detect, diagnose and classify faults in edge computing environments. PreGAN uses GAT and GRU for feature extraction with a Multi-Head-Attention and Prototype prediction decoder to detect and classify faults. PreGAN leverages a generator model to utilize the anomaly class prototypes to output a delta scheduling decision (migrations) to rectify the faults and improve QoS. The discriminator model with co-simulations allow PreGAN to decide between the original and modified scheduling decisions. Moreover, PreGAN's discriminator aids generator training using adversarial loss and does not require it to run co-simulations at test time. This allows PreGAN to have high detection and classification accuracies that aid efficient fault recovery for optimal QoS. Specifically, PreGAN achieves an improvement of 8\%, 5\% and 12\% for energy consumption, response times and SLO violations respectively. It is also able to correctly identify faults, giving an average F1 score of 0.8868, higher than the state-of-the-art models. PreGAN is able to achieve this with 23.8\% lower overheads compared to the baseline method with the highest F1 score. This makes PreGAN an ideal choice for reliable edge computing with time-critical applications.

We now present some future directions for this work. Due to difficulty in obtaining labeled data for training, we propose to extend PreGAN to utilize unsupervised models~\cite{topomad}. Finally, the current model assumes a master-slave design and we plan to explore extensions that allow us to deploy PreGAN in federated or serverless platforms with streaming tasks~\cite{casale2020radon}. 

\section*{Software Availability}
The PreGAN code is available at \url{https://github.com/imperial-qore/PreGAN} under BSD-3 License. The Docker images are available at \url{https://hub.docker.com/u/shreshthtuli}.

\section*{Acknowledgments}
Shreshth Tuli is supported by the President’s Ph.D. Scholarship at Imperial College London. 

\bibliographystyle{IEEEtran}
\bibliography{references}

\begin{thebibliography}{10}
\providecommand{\url}[1]{#1}
\csname url@samestyle\endcsname
\providecommand{\newblock}{\relax}
\providecommand{\bibinfo}[2]{#2}
\providecommand{\BIBentrySTDinterwordspacing}{\spaceskip=0pt\relax}
\providecommand{\BIBentryALTinterwordstretchfactor}{4}
\providecommand{\BIBentryALTinterwordspacing}{\spaceskip=\fontdimen2\font plus
\BIBentryALTinterwordstretchfactor\fontdimen3\font minus
  \fontdimen4\font\relax}
\providecommand{\BIBforeignlanguage}[2]{{%
\expandafter\ifx\csname l@#1\endcsname\relax
\typeout{** WARNING: IEEEtran.bst: No hyphenation pattern has been}%
\typeout{** loaded for the language `#1'. Using the pattern for}%
\typeout{** the default language instead.}%
\else
\language=\csname l@#1\endcsname
\fi
#2}}
\providecommand{\BIBdecl}{\relax}
\BIBdecl

\bibitem{gill2019transformative}
S.~S. Gill, S.~Tuli, M.~Xu, I.~Singh, K.~V. Singh, D.~Lindsay, S.~Tuli,
  D.~Smirnova, M.~Singh, U.~Jain \emph{et~al.}, ``{Transformative effects of
  IoT, Blockchain and Artificial Intelligence on cloud computing: Evolution,
  vision, trends and open challenges},'' \emph{Internet of Things}, vol.~8, pp.
  100--118, 2019.

\bibitem{tuli2020dynamic}
S.~Tuli, S.~Ilager, K.~Ramamohanarao, and R.~Buyya, ``{Dynamic Scheduling for
  Stochastic Edge-Cloud Computing Environments using A3C learning and Residual
  Recurrent Neural Networks},'' \emph{IEEE Transactions on Mobile Computing},
  2020.

\bibitem{pcft}
J.~Liu, S.~Wang, A.~Zhou, S.~A. Kumar, F.~Yang, and R.~Buyya, ``Using proactive
  fault-tolerance approach to enhance cloud service reliability,'' \emph{IEEE
  Transactions on Cloud Computing}, vol.~6, no.~4, pp. 1191--1202, 2016.

\bibitem{dastjerdi2016fog}
A.~V. Dastjerdi and R.~Buyya, ``Fog computing: Helping the internet of things
  realize its potential,'' \emph{Computer}, vol.~49, no.~8, pp. 112--116, 2016.

\bibitem{nicoletti2013cloud}
B.~Nicoletti, \emph{Cloud computing in financial services}.\hskip 1em plus
  0.5em minus 0.4em\relax Springer, 2013.

\bibitem{park2018lired}
D.~Park, S.~Kim, Y.~An, and J.-Y. Jung, ``{LiReD: A light-weight real-time
  fault detection system for edge computing using LSTM recurrent neural
  networks},'' \emph{Sensors}, vol.~18, no.~7, p. 2110, 2018.

\bibitem{malik2011adaptive}
S.~Malik and F.~Huet, ``Adaptive fault tolerance in real time cloud
  computing,'' in \emph{2011 IEEE World Congress on services}.\hskip 1em plus
  0.5em minus 0.4em\relax IEEE, 2011, pp. 280--287.

\bibitem{ristov2020resilient}
S.~Ristov, T.~Fahringer, D.~Peer, T.-P. Pham, M.~Gusev, and C.~Mas-Machuca,
  ``Resilient techniques against disruptions of volatile cloud resources,'' in
  \emph{Guide to Disaster-Resilient Communication Networks}.\hskip 1em plus
  0.5em minus 0.4em\relax Springer, 2020, pp. 379--400.

\bibitem{eclb}
A.~Sharif, M.~Nickray, and A.~Shahidinejad, ``{Fault-tolerant with load
  balancing scheduling in a fog-based IoT application},'' \emph{IET
  Communications}, vol.~14, no.~16, pp. 2646--2657, 2020.

\bibitem{cmodlb}
S.~Negi, M.~M.~S. Rauthan, K.~S. Vaisla, and N.~Panwar, ``{CMODLB: an efficient
  load balancing approach in cloud computing environment},'' \emph{The Journal
  of Supercomputing}, pp. 1--53, 2021.

\bibitem{dftm}
V.~Sivagami and K.~Easwarakumar, ``An improved dynamic fault tolerant
  management algorithm during vm migration in cloud data center,'' \emph{Future
  Generation Computer Systems}, vol.~98, pp. 35--43, 2019.

\bibitem{kumari2018survey}
P.~Kumari and P.~Kaur, ``A survey of fault tolerance in cloud computing,''
  \emph{Journal of King Saud University-Computer and Information Sciences},
  2018.

\bibitem{zhou2010security}
M.~Zhou, R.~Zhang, W.~Xie, W.~Qian, and A.~Zhou, ``Security and privacy in
  cloud computing: A survey,'' in \emph{2010 Sixth International Conference on
  Semantics, Knowledge and Grids}.\hskip 1em plus 0.5em minus 0.4em\relax IEEE,
  2010, pp. 105--112.

\bibitem{zheng2011component}
Z.~Zheng, T.~C. Zhou, M.~R. Lyu, and I.~King, ``Component ranking for
  fault-tolerant cloud applications,'' \emph{IEEE Transactions on Services
  Computing}, vol.~5, no.~4, pp. 540--550, 2011.

\bibitem{hong2019resource}
C.-H. Hong and B.~Varghese, ``Resource management in fog/edge computing: a
  survey on architectures, infrastructure, and algorithms,'' \emph{ACM
  Computing Surveys (CSUR)}, vol.~52, no.~5, pp. 1--37, 2019.

\bibitem{khan2019edge}
W.~Z. Khan, E.~Ahmed, S.~Hakak, I.~Yaqoob, and A.~Ahmed, ``Edge computing: A
  survey,'' \emph{Future Generation Computer Systems}, vol.~97, pp. 219--235,
  2019.

\bibitem{engelmann2009proactive}
C.~Engelmann, G.~R. Vallee, T.~Naughton, and S.~L. Scott, ``Proactive fault
  tolerance using preemptive migration,'' in \emph{2009 17th Euromicro
  International Conference on Parallel, Distributed and Network-based
  Processing}.\hskip 1em plus 0.5em minus 0.4em\relax IEEE, 2009, pp. 252--257.

\bibitem{tian2018scheduling}
B.~Tian, C.~Tian, H.~Dai, and B.~Wang, ``Scheduling coflows of multi-stage jobs
  to minimize the total weighted job completion time,'' in \emph{IEEE INFOCOM
  2018-IEEE Conference on Computer Communications}.\hskip 1em plus 0.5em minus
  0.4em\relax IEEE, 2018, pp. 864--872.

\bibitem{mad_gan}
D.~Li, D.~Chen, B.~Jin, L.~Shi, J.~Goh, and S.-K. Ng, ``{MAD-GAN: Multivariate
  anomaly detection for time series data with generative adversarial
  networks},'' in \emph{International Conference on Artificial Neural
  Networks}.\hskip 1em plus 0.5em minus 0.4em\relax Springer, 2019, pp.
  703--716.

\bibitem{gat}
P.~Veli{\v{c}}kovi{\'c}, G.~Cucurull, A.~Casanova, A.~Romero, P.~Lio, and
  Y.~Bengio, ``Graph attention networks,'' \emph{arXiv preprint
  arXiv:1710.10903}, 2017.

\bibitem{gru}
J.~Chung, C.~Gulcehre, K.~Cho, and Y.~Bengio, ``Empirical evaluation of gated
  recurrent neural networks on sequence modeling,'' in \emph{NIPS 2014 Workshop
  on Deep Learning, December 2014}, 2014.

\bibitem{vaswani2017attention}
A.~Vaswani, N.~Shazeer, N.~Parmar, J.~Uszkoreit, L.~Jones, A.~N. Gomez,
  {\L}.~Kaiser, and I.~Polosukhin, ``Attention is all you need,'' in
  \emph{Proceedings of the 31st International Conference on Neural Information
  Processing Systems}, 2017, pp. 6000--6010.

\bibitem{snell2017prototypical}
J.~Snell, K.~Swersky, and R.~Zemel, ``Prototypical networks for few-shot
  learning,'' in \emph{Proceedings of the 31st International Conference on
  Neural Information Processing Systems}, 2017, pp. 4080--4090.

\bibitem{tuli2021cosco}
S.~Tuli, S.~R. Poojara, S.~N. Srirama, G.~Casale, and N.~R. Jennings, ``{COSCO:
  Container Orchestration Using Co-Simulation and Gradient Based Optimization
  for Fog Computing Environments},'' \emph{IEEE Transactions on Parallel and
  Distributed Systems}, vol.~33, no.~1, pp. 101--116, 2022.

\bibitem{ataallah2015fault}
S.~M. Ataallah, S.~M. Nassar, and E.~E. Hemayed, ``Fault tolerance in cloud
  computing-survey,'' in \emph{2015 11th International computer engineering
  conference (ICENCO)}.\hskip 1em plus 0.5em minus 0.4em\relax IEEE, 2015, pp.
  241--245.

\bibitem{pbfm}
B.~Ray, A.~Saha, S.~Khatua, and S.~Roy, ``Proactive fault-tolerance technique
  to enhance reliability of cloud service in cloud federation environment,''
  \emph{IEEE Transactions on Cloud Computing}, 2020.

\bibitem{mohammed2017failover}
B.~Mohammed, M.~Kiran, K.~M. Maiyama, M.~M. Kamala, and I.-U. Awan, ``Failover
  strategy for fault tolerance in cloud computing environment,''
  \emph{Software: Practice and Experience}, vol.~47, no.~9, pp. 1243--1274,
  2017.

\bibitem{satpathy2018crow}
A.~Satpathy, S.~K. Addya, A.~K. Turuk, B.~Majhi, and G.~Sahoo, ``Crow search
  based virtual machine placement strategy in cloud data centers with live
  migration,'' \emph{Computers \& Electrical Engineering}, vol.~69, pp.
  334--350, 2018.

\bibitem{wang2021ddqp}
L.~Wang, W.~Mao, J.~Zhao, and Y.~Xu, ``{DDQP: A double deep Q-learning approach
  to online fault-tolerant SFC placement},'' \emph{IEEE Transactions on Network
  and Service Management}, vol.~18, no.~1, pp. 118--132, 2021.

\bibitem{tuli2019fogbus}
S.~Tuli, R.~Mahmud, S.~Tuli, and R.~Buyya, ``Fogbus: A blockchain-based
  lightweight framework for edge and fog computing,'' \emph{Journal of Systems
  and Software}, 2019.

\bibitem{basu2019learn}
D.~Basu, X.~Wang, Y.~Hong, H.~Chen, and S.~Bressan, ``Learn-as-you-go with
  megh: Efficient live migration of virtual machines,'' \emph{IEEE Transactions
  on Parallel and Distributed Systems}, vol.~30, no.~8, pp. 1786--1801, 2019.

\bibitem{audibert2020usad}
J.~Audibert, P.~Michiardi, F.~Guyard, S.~Marti, and M.~A. Zuluaga, ``{USAD:
  UnSupervised Anomaly Detection on Multivariate Time Series},'' in
  \emph{Proceedings of the 26th ACM SIGKDD International Conference on
  Knowledge Discovery \& Data Mining}, 2020, pp. 3395--3404.

\bibitem{tuli2021tango}
S.~Tuli, R.~Bansal, R.~Paul \emph{et~al.}, ``{TANGO: Commonsense Generalization
  in Predicting Tool Interactions for Mobile Manipulators},''
  \emph{International Joint Conference on Artificial Intelligence (IJCAI)},
  2021.

\bibitem{tuli2021hunter}
S.~Tuli, S.~S. Gill, M.~Xu, P.~Garraghan, R.~Bahsoon, S.~Dustdar,
  R.~Sakellariou, O.~Rana, R.~Buyya, G.~Casale \emph{et~al.}, ``{HUNTER: AI
  based holistic resource management for sustainable cloud computing},''
  \emph{Journal of Systems and Software}, pp. 111--124, 2021.

\bibitem{paszke2019pytorch}
A.~Paszke, S.~Gross, F.~Massa, A.~Lerer, J.~Bradbury, G.~Chanan, T.~Killeen,
  Z.~Lin, N.~Gimelshein, L.~Antiga \emph{et~al.}, ``{PyTorch: An Imperative
  Style, High-Performance Deep Learning Library},'' \emph{Advances in Neural
  Information Processing Systems}, vol.~32, pp. 8026--8037, 2019.

\bibitem{spec}
\BIBentryALTinterwordspacing
{Standard Performance Evaluation Corporation}. Spec power consumption models.
  [Online]. Available: \url{https://www.spec.org/cloud_iaas2018/results/}
\BIBentrySTDinterwordspacing

\bibitem{mcchesney2019defog}
J.~McChesney, N.~Wang, A.~Tanwer, E.~de~Lara, and B.~Varghese, ``{DeFog: fog
  computing benchmarks},'' in \emph{The 4th ACM/IEEE Symposium on Edge
  Computing}, 2019, pp. 47--58.

\bibitem{mao2016dynamic}
Y.~Mao, J.~Zhang, and K.~B. Letaief, ``Dynamic computation offloading for
  mobile-edge computing with energy harvesting devices,'' \emph{IEEE Journal on
  Selected Areas in Communications}, vol.~34, no.~12, pp. 3590--3605, 2016.

\bibitem{mtad_gat}
H.~Zhao, Y.~Wang, J.~Duan, C.~Huang, D.~Cao, Y.~Tong, B.~Xu, J.~Bai, J.~Tong,
  and Q.~Zhang, ``Multivariate time-series anomaly detection via graph
  attention network,'' \emph{International Conference on Data Mining}, 2020.

\bibitem{omnianomaly}
Y.~Su, Y.~Zhao, C.~Niu, R.~Liu, W.~Sun, and D.~Pei, ``Robust anomaly detection
  for multivariate time series through stochastic recurrent neural network,''
  in \emph{Proceedings of the 25th ACM SIGKDD International Conference on
  Knowledge Discovery \& Data Mining}, 2019, pp. 2828--2837.

\bibitem{jarvelin2002cumulated}
K.~J{\"a}rvelin and J.~Kek{\"a}l{\"a}inen, ``Cumulated gain-based evaluation of
  ir techniques,'' \emph{ACM Transactions on Information Systems (TOIS)},
  vol.~20, no.~4, pp. 422--446, 2002.

\bibitem{ade}
\BIBentryALTinterwordspacing
{Open Mainframe Project}. {Anomaly Detection Engine for Linux Logs (ADE)}.
  [Online]. Available:
  \url{https://www.openmainframeproject.org/projects/anomaly-detection-engine-for-linux-logs-ade}
\BIBentrySTDinterwordspacing

\bibitem{saleh2019dynamic}
N.~Saleh and M.~Mashaly, ``A dynamic simulation environment for container-based
  cloud data centers using containercloudsim,'' in \emph{2019 Ninth
  International Conference on Intelligent Computing and Information Systems
  (ICICIS)}.\hskip 1em plus 0.5em minus 0.4em\relax IEEE, 2019, pp. 332--336.

\bibitem{beloglazov2012optimal}
A.~Beloglazov and R.~Buyya, ``Optimal online deterministic algorithms and
  adaptive heuristics for energy and performance efficient dynamic
  consolidation of virtual machines in cloud data centers,'' \emph{Concurrency
  and Computation: Practice and Experience}, vol.~24, no.~13, pp. 1397--1420,
  2012.

\bibitem{topomad}
Z.~He, P.~Chen, X.~Li, Y.~Wang, G.~Yu, C.~Chen, X.~Li, and Z.~Zheng, ``A
  spatiotemporal deep learning approach for unsupervised anomaly detection in
  cloud systems,'' \emph{IEEE Transactions on Neural Networks and Learning
  Systems}, 2020.

\bibitem{casale2020radon}
G.~Casale, M.~Arta{\v{c}}, W.-J. van~den Heuvel, A.~van Hoorn, P.~Jakovits,
  F.~Leymann, M.~Long, V.~Papanikolaou, D.~Presenza, A.~Russo \emph{et~al.},
  ``Radon: rational decomposition and orchestration for serverless computing,''
  \emph{SICS Software-Intensive Cyber-Physical Systems}, vol.~35, no.~1, pp.
  77--87, 2020.

\end{thebibliography}

\end{document}